\documentstyle[12pt]{article}
\def \beq{\begin{equation}}
\def \eeq{
\end{equation}}
\def \s{\sqrt{2}}
\def\ubar{\overline{u}}

\def\dbar{\overline{d}}
\def\sbar{\overline{s}}
\def\bbar{\overline{b}}
\def\fbar{\overline{f}}
\def\qbar{\overline{q}}

\def\Abar{\overline{A}}
\def\Kbar{\overline{K}}
\def\Bbar{\overline{B}^0}

\def\bd{B^0_d}
\def\bs{B_s^0}
\def\bdb{\overline{B}^0_d}
\def\bsb{\overline{B}_s^0}

\def\Dbar{\overline{D}^0}

\def\Gammabar{\overline{\Gamma}}
\def\to{\rightarrow}
\begin{document}
\begin{titlepage}
\rightline{TECHNION-PH-96-39}
\rightline{hep-ph/9609430}
\rightline{September 1996}
\bigskip
\bigskip

\large
\centerline {\bf CP Violation in B-Decays:}
\centerline {\bf The Standard Model and Beyond~\footnote
{Invited talk given at Beauty 96, Rome, June 17$-$21, 1996, to appear
in the proceedings}}
\normalsize
\vskip 2.0cm
\centerline {Michael Gronau}
\centerline {\it Department of Physics}
\centerline {\it Technion - Israel Institute of Technology, 32000 Haifa, Israel}
\vskip 4.0cm

\centerline {\bf Abstract}
\vskip 1.0cm
We review the subject of CP violation in $B$ decays in the Standard Model (SM)
and  beyond the SM. We describe some of the present most promising ways of
testing the Cabibbo-Kobayashi-Maskawa (CKM) origin of CP violation through
a determination of the three angles of the CKM matrix unitarity triangle.
Other sources of CP nonconservation violate SM constraints on the unitarity
triangle.
We show that different models of physics beyond the SM can be distinguished by
combining their effects in CP asymmetries and in rare flavor-changing $B$
decays.

\vfill
\end{titlepage}

\newpage
\section{Introduction}

One of the remaining goals in the present era of particle physics is to study
the origin of CP violation. In the Standard Model (SM), CP nonconservation is
due to a nonzero complex phase in the Cabibbo-Kobayashi-Maskawa (CKM) matrix
$V$, describing the interaction of the three families of left-handed
quarks with the charged gauge boson. This unitary matrix can be approximated
by the following two useful forms \cite{PDG}:
$$
V \approx \left(\matrix{1-{1\over 2}s_{12}^2&s_{12}&s_{13}e^{-i\gamma}\cr
-s_{12}&1-{1\over 2}s_{12}^2&s_{23}\cr
s_{12}s_{23}-s_{13}e^{i\gamma}&-s_{23}&1\cr}\right)
$$
\beq \label{V}
\approx \left(\matrix{1-{1\over 2}\lambda^2&\lambda&A\lambda^3(\rho-i\eta)\cr
 -\lambda&1-{1\over 2}\lambda^2&A\lambda^2\cr
A\lambda^3(1-\rho-i\eta)&-A\lambda^2&1\cr}\right)~.
\eeq

The measured values of the three Euler-like mixing angles , $\theta_{ij}$, and
the phase $\gamma$ are given by \cite{ALON}:
$$
s_{12}\equiv \sin\theta_{12}\approx\vert V_{us}\vert=0.220\pm 0.002~,
$$
$$
s_{23}\equiv \sin\theta_{23}\approx\vert V_{cb}\vert=0.039\pm 0.003~,
$$
$$
~~~~s_{13}\equiv \sin\theta_{13}\equiv\vert V_{ub}\vert=0.0031\pm 0.0009~,
$$
\beq \label{ANGLES}
35^0\leq\gamma\equiv{\rm Arg}(V^*_{ub})\leq 145^0~.
\eeq
The only information about a nonzero value of $\gamma$
comes from CP violation in the $K^0-\Kbar$ system.

Unitarity of $V$ implies triangle relations such as
\beq\label{UNIT}
V_{ud}V^*_{ub}+V_{cd}V^*_{cb}+V_{td}V^*_{tb}=0~,
\eeq
which is shown in Fig.~1. The angle $\alpha$ of the unitarity triangle
has rather crude bounds, qualitatively similar to those of $\gamma$,
\beq\label{ALPHA}
20^0\leq\alpha\leq 120^0~,
\eeq
whereas $\beta$ is somewhat better determined
\beq\label{BETA}
10^0\leq\beta\leq 35^0~.
\eeq
In addition to the constraints on $\alpha,~\beta$ and $\gamma$, pairs of
these angles are correlated. Due to the rather limited range of $\beta$, the
angles $\alpha$ and $\gamma$ are almost linearly correlated through ~$\alpha +
\gamma = \pi- \beta$ \cite{DGR}.
A special correlation exists also between small values of $\sin 2\beta$ and
large values of $\sin 2\alpha$ \cite{NISAR}.

A precise determination of the three angles $\alpha,~\beta$ and $\gamma$,
which would provide a test of the CKM origin of CP violation, depends
crucially on measuring CP asymmetries in $B$ decays. This will be one
of the central subjects of this paper. Since much of this material was
reviewed
in last year {\it Beauty 95}, I will be quite brief on most
topics. The reader is referred to ref. \cite{B95} for details and
further references. I will rather concentrate on new methods which were
developed since last year. The second subject of this paper will be
manifestations of physics beyond the standard model in CP asymmetries, on the
one hand, and in rare flavor-changing $B$-decays, on the other hand.

Section 2 is a quick run through the by-now standard methods of measuring
$\alpha,~\beta$ and
$\gamma$ in neutral and charged $B$ decays. The role of isospin symmetry in
resolving ``penguin pollution" is briefly discussed. Section 3 studies
flavor SU(3) and first-order SU(3) breaking in $B$ decays to two pseudoscalar
mesons. The power of this analysis is demonstrated by an example in which both
$\alpha$ and $\gamma$ can be measured in $B$ decays to kaons and charged pions.
The use of $\eta$ and $\eta'$ in final states is briefly discussed. Section 4
reviews CP asymmetries beyond the SM, while Section 5 concludes.

\section{"Standard" methods of measuring $\alpha,~\beta,~\gamma$}

\subsection{Neutral $B$ Decays to CP-eigenstates}

The most frequently discussed method of measuring weak phases is based on
neutral
$B$ decays to final states $f$ which are common to $B^0$ and $\Bbar$. CP
violation is induced by $B^0-\Bbar$ mixing through the interference of the two
amplitudes $B^0\to f$ and $B^0\to\Bbar\to f$. When $f$ is a CP-eigenstate, and
when a single weak amplitude (or rather a single weak phase) dominates the
decay process, the time-dependent asymmetry
\beq
{\cal A}(t)\equiv {\Gamma(B^0(t)\to f)-\Gamma(\Bbar (t)\to f)\over
\Gamma(B^0(t)\to f)+\Gamma(\Bbar (t)\to f)}
\eeq
obtains the simple form \cite{BIGSAN}
\beq\label{ASYM}
{\cal A}(t)= \xi\sin2(\phi_M+\phi_f)\sin(\Delta mt)~.
\eeq
$\xi$ is the CP eigenvalue of $f$, $2\phi_M$ is the phase of $B^0-\Bbar$ mixing,
($\phi_M=\beta,~0$ for $B^0_d,~B^0_s$, respectively), $\phi_f$ is the weak
phase of the $B^0\to f$ amplitude, and $\Delta m$ is the neutral $B$
mass-difference.

The two very familar examples are:

(i) $B^0_d\to \psi K_S$, where $\xi=-1,~\phi_f={\rm Arg}(V^*_{cb}V_{cs})=0$,
\beq
{\cal A}(t)=-\sin2\beta\sin(\Delta mt)~,
\eeq
and

(ii) $B^0_d\to\pi^+\pi^-$, where $\xi=1,~\phi_f={\rm Arg}(V^*_{ub}V_{ud})=
\gamma$,
\beq
{\cal A}(t)=-\sin2\alpha\sin(\Delta mt)~.
\eeq
Thus, the two asymmetries measure the angles $\beta$ and $\alpha$.

\subsection{Decays to other states}

A similar method can also be applied to measure weak phases when $f$ is a
common decay  mode of $B^0$ and $\Bbar$, but not
necessarily a CP eigenstate. In this case one measures four different
time-dependent decay rates, $\Gamma_f(t),~\Gammabar_f(t),~\Gamma_{\fbar}(t),~
\Gammabar_{\fbar}(t)$, corresponding to initial $B^0$ and $\Bbar$ decaying to
$f$ and its charge-conjugate $\fbar$ \cite{MG}. The four rates depend on four
unknown quantities, $|A|,~|\Abar|,~\sin(\Delta\delta_f+\Delta\phi_f+2\phi_M),~
\sin(\Delta\delta_f-\Delta\phi_f-2\phi_M)$. ($A$ and $\Abar$ are the decay
amplitudes of $B^0$ and $\Bbar$ to $f$, $\Delta\delta_f$
and $\Delta\phi_f$ are the the strong and weak phase-differences between these
amplitudes). Thus, the four rate measurements allow a determination of the weak
CKM phase $\Delta\phi_f+2\phi_M$. This method can be applied to measure
$\alpha$ in $B^0_d\to\rho^+\pi^-$, and to measure $\gamma$ in
$B^0_s\to D^+_s K^-$ \cite{ADKD}. Other ways of measuring $\gamma$ in $B^0_s$
decays were discussed recently in Ref.~\cite{FLEIDU}.

\subsection{``Penguin pollution"}

All this assumes that a single weak phase dominates the decay
$B^0(\Bbar)\to f$. As a matter of fact, in a variety of decay processes, such
as in $\bd\to\pi^+\pi^-$, there exists a second amplitude due to a ``penguin"
diagram in addition to the usual ``tree" diagram \cite{PEN}. As a result, CP is
also violated in the direct decay of a $B^0$, and one faces a problem of
separating the two types of asymmetries. This can only be partially
achieved through the more general time-dependence
\beq\label{GENASYM}
{\cal A}(t)=
{(1-|\Abar/A|^2)\cos(\Delta mt)-2{\rm
Im}(e^{-2i\phi_M}\Abar/A)\sin(\Delta mt)\over 1+|\Abar/A|^2}~.
\eeq
Here the $\cos(\Delta mt)$ term implies direct CP violation, and the
coefficient of $\sin(\Delta mt)$ obtains a correction from the penguin
amplitude. The two terms have a different dependence on $\Delta\delta$,
the final-state phase-difference between the tree and penguin amplitudes.
The coefficient of $\cos(\Delta mt)$ is proportional to $\sin(\Delta\delta)$,
whereas the correction to the coefficient of $\sin(\Delta mt)$ is proportional
to $\cos(\Delta\delta)$. Thus, if $\Delta\delta$ were small, this correction
might be large in spite of the fact that the $\cos(\Delta mt)$ term were too
small to be observed.

\subsection{Isospin resolution of penguin pollution}

The above ``penguin pollution" may lead to dangerously large effects in
$B^0_d(t)\to\pi^+\pi^-$ decay, which would avoid a clean determination of
$\alpha$ \cite{PENPI}. One way to remove this effect is by measuring also the
(time-integrated) rates of $\bd\to\pi^0\pi^0$, $B^+\to\pi^+\pi^0$
and their charge-conjugates \cite{GRLO}. One uses the different isospin
properties
of the penguin ($\Delta I=1/2$) and tree ($\Delta I=1/2, 3/2$) operators and
the well-defined weak phase of the tree operator. This enables one to determine
the correction to $\sin2\alpha$ in the second term of Eq.(\ref{GENASYM}).
Electroweak penguin contributions could, in principle, spoil this method, since
unlike the QCD penguins they are not pure $\Delta I=1/2$ \cite{DH}. These
effects
are, however, very small and consequently lead to a tiny uncertainty in
determining $\alpha$ \cite{EWP}. The difficult part of this method
seems to be the necessary observation of the decay to two neutral pions
which is expected to be color-suppressed. Other methods of resolving the
``penguin pollution" in $\bd\to\pi^+\pi^-$, which do not rely on decays
to neutral pions, will be described in Sec. 3.

\subsection{Measuring $\gamma$ in $B^{\pm}\to D K^{\pm}$}

In $B^{\pm}\to D K^{\pm}$, where $D$ may be either a flavor state ($D^0,~
\Dbar$) or a CP-eigenstate ($D^0_1,~D^0_2$), one can measure separately the
magnitudes of two interfering amplitudes leading to direct CP violation.
This enables a measurement of $\gamma$, the relative weak phase between these
two amplitudes \cite{GW}. This method is based on a simple quantum mechanical
relation among the amplitudes of three different processes,
\beq
\sqrt{2}A(B^+\to D^0_1 K^+)~=~A(B^+\to D^0 K^+)~+~A(B^+\to \Dbar K^+)~.
\eeq
The CKM factors of the two terms on the right-hand-side,
$V^*_{ub}V^{~}_{cs}$ and $V^*_{cb}V^{~}_{us}$, involve the weak phases $\gamma$
and zero, respectively. A similar triangle relation can be written for the
charge-conjugate processes. Measurement of the rates of these six proccesses,
two pairs of which are equal, enables a determination of $\gamma$.
The present upper limit on the branching ratio of $B^+\to \overline{D}^0 K^+$
\cite{DK} is already very close to the value expected in the SM. The major
difficulty of this method may be in measuring $B^+\to D^0 K^+$ which is expected
to be color-suppressed. For further details and a feasibility study see
Ref.~\cite{STONE}.

\section{Flavor SU(3) and SU(3) breaking}

\subsection{The general formalism}

One may use approximate flavor SU(3) symmetry of strong interactions to relate
all two body processes of the type $B\to\pi\pi,~B\to\pi K$ and $B\to K\Kbar$.
Since SU(3) is expected to be broken by effects of order $20\%$, such
as in $f_K/f_{\pi}$, one must introduce SU(3) breaking terms in such an
analysis.
This approach has recently received special attention
\cite{GHLR,WOLF,DHE,BF,KP,GL}.
In the present section we will discuss two applications of this analysis to
a determination of weak phases.
Early applications of SU(3) to two-body $B$ decays can be found in
Ref.~\cite{EARLY}.

The weak Hamiltonian operators associated with the transitions
$\bbar\to\ubar uq$
and $\bbar\to \qbar$ ($q=d$ or $s$) transform as a ${\bf 3^*},~{\bf 6}$ and
${\bf 15^*}$ of SU(3). The $B$ mesons are in a triplet, and the symmetric
product of two final state pseudoscalar octets in an S-wave contains a singlet,
an octet and a 27-plet. Thus, all these processes can be described in terms of
five SU(3) amplitudes: $\langle  ~1~ || ~3^* || 3 \rangle,~\langle  8 ||
~3^* || 3
\rangle,~\langle  8 || ~6~  || 3 \rangle,~\langle  8 || 15^* || 3 \rangle,~
\langle 27 || 15^* || 3 \rangle$.

An equivalent and considerably more convenient representation of these
amplitudes is given in terms of an overcomplete set of six quark diagrams
occuring in five different combinations.
These diagrams are denoted by $T$ (tree), $C$ (color-suppressed),
$P$ (QCD-penguin), $E$ (exchange), $A$ (annihilation) and $PA$ (penguin
annihilation). The last three amplitudes, in which the spectator quark enters
into the decay Hamiltonian, are expected to be suppressed by $f_B/m_B$
($f_B\approx 180~{\rm MeV}$) and may be neglected to a good approximation.

The presence of higher-order electroweak penguin contributions introduces no new
SU(3) amplitudes, and in terms of quark graphs merely leads to a substitution
\cite{EWP}
\beq \label{eqn:combs}
T\to t\equiv T + P^C_{EW}~~,~~
C\to c\equiv C + P_{EW}~~,~~
P\to p\equiv P-{1\over 3}P^C_{EW}~~,
\eeq
where $P_{EW}$ and $P^C_{EW}$ are color-favored and color-suppressed
electroweak penguin amplitudes. $\Delta S=0$ amplitudes are denoted by unprimed
quantities and $|\Delta S|=1$ processes by primed quantities. Corresponding
ratios are given by ratios of CKM factors
\beq
{T'\over T} = {C'\over C} = {V_{us}\over V_{ud}}~,~~~~~~{P'\over P} =
{P'_{EW}\over P_{EW}} = {V_{ts}\over V_{td}}~.
\eeq
$t$-dominance was assumed in the ratio $P'/P$. The effect of $u$ and $c$ quarks
in penguin amplitudes can sometimes be important \cite{BUFLE}.

The expressions of all two body decays to two light pseudoscalars in the
SU(3) limit are given in Table 1.
The vanishing of three of the amplitudes, associated with $B^0_d\to K^+ K^-,~
B^0_s\to\pi^+\pi^-,~B^0_s\to\pi^0\pi^0$, follows from the assumption of
negligible exchange ($E$) amplitudes. This can be used to test our assumption
which neglects final state rescattering effects.

First-order SU(3) breaking corrections can be introduced in a most general
manner through parameters describing mass insertions in the above quark
diagrams \cite{SU3BR}. The interpretation of these corrections in terms of
ratios of decay constants and form factors is model-dependent. There is,
however, one  case in which such interpretation is quite reliable. Consider
the tree amplitudes $T$ and $T'$. In $T$ the $W$ turns into a $u\dbar$ pair,
whereas in $T'$ it turns into $u\sbar$. One may assume factorization for
$T$ and
$T'$, which is supported by data on $B\to \overline{D}\pi$ \cite{BS}, and is
justified for $B\to\pi\pi$ and $B\to\pi K$ by the high momentum with which
the two color-singlet mesons separate from one another. Thus,
\beq\label{SU3BRK}
{T'\over T} = {V_{us}\over V_{ud}}{f_K\over f_{\pi}}~.
\eeq
Similar assumptions for $C'/C$ and $P'/P$ cannot  be justified.

\subsection{$\alpha$ and $\gamma$ from $B$ decays to kaons and charged pions}

Table 1 and Eq.(\ref{SU3BRK}) can be used to separate the penguin term from
the tree amplitude in $B^0_d\to\pi^+\pi^-$, and thereby determine
simultaneously
both the angles $\alpha$ and $\gamma$. In the present subsection we
outline in a schematic way a method which, for a practical purpose, uses only
final states with kaons and charged pions. The reader is
referred to Ref.~\cite{MGJR} for more details. A few alternative ways to learn
the penguin effects in $B^0_d\to\pi^+\pi^-$ were suggested in Ref.~\cite{ALT}

Consider the amplitudes of the three processes $B^0_d\to\pi^+\pi^-,
~B^0_d\to\pi^- K^+,~B^+\to\pi^+ K^0$ given by
\beq
A(B^0_d\to\pi^+\pi^-) = -t-p,~A(B^0_d\to\pi^- K^+)=-t'-p',~
A(B^+\to\pi^+ K^0)=p'~.
\eeq
One measures the time-dependence of the first process and the decay rates of
the other two self-tagging modes. Using these
measurements, and the ones corresponding to the charge-conjugate processes,
one can form the following six measurables:
$$
~\Gamma(B^0_d(t)\to\pi^+\pi^-) + \Gamma(\bdb(t)\to\pi^+\pi^-) =
{\bf A}e^{-t/\tau_B}
$$
$$
~~~~~~~~\Gamma(B^0_d(t)\to\pi^+\pi^-) - \Gamma(\bdb(t)\to\pi^+\pi^-) =
[{\bf B}\cos(\Delta mt)
$$
$$
~~~~~~~~~~~~~~~~~~~~~~~~~~~~~~~~~~~~~
~~~~~~~~~~~~~~~~~~~~~~~~~~~~~~~+ {\bf C}\sin(\Delta mt)]e^{-t/\tau_B}
$$
$$
\Gamma(\bd\to \pi^- K^+) + \Gamma(\bdb\to \pi^+ K^-) = {\bf D}
$$
$$
\Gamma(\bd\to \pi^- K^+) - \Gamma(\bdb\to \pi^+ K^-) = {\bf E}
$$
\beq
\Gamma(B^+\to \pi^+ K^0) = \Gamma(B^-\to \pi^- \overline{K}^0) = {\bf F}
\eeq

The six quantities ${\bf A, B,...F}$ can all be expressed in terms
six parameters, consisting of the magnitudes and the weak and strong phases of
the tree and penguin amplitudes. Let's count these parameters:
\begin{itemize}
\item The $\Delta S=0$ tree amplitude $T$ has magnitude ${\cal T}$, weak phase
$\gamma$ and strong phase $\delta_T$.
\item The $\Delta S=0$ penguin amplitude $P+(2/3)P^C_{EW}$ has magnitude
${\cal P}$, weak phase $-\beta$ and strong phase $\delta_P$.
\item The $\Delta S=1$ tree amplitude $T'$ has magnitude  $(V_{us}/V_{ud})
(f_K/f_{\pi}){\cal T}$, weak phase $\gamma$ and strong phase $\delta_T$.
\item The $\Delta S=1$ penguin amplitude $P'+(2/3)P'^C_{EW}$ has magnitude
${\cal P'}$, weak phase $\pi$ and strong phase $\delta_P$. (Since the strong
phase difference $\delta_T-\delta_P$ is expected to be small, we neglect SU(3)
breaking effects in this phase).
\end{itemize}
Therefore, the six measurables ${\bf A, B,...F}$ can be  expressed in terms of
the six parameters ${\cal T},~{\cal P},~{\cal
P'},~\delta\equiv\delta_T-\delta_P,
~\gamma$ and $\alpha\equiv\pi-\beta-\gamma$. This enables a determination
of both
$\alpha$ and $\gamma$, with some remaining discrete ambiguity associated with
the size of final-state phases. A sample of events corresponding to about
100 $B^0_d\to\pi^+\pi^-$, 100 $B^0_d\to\pi^{\pm}K^{\mp}$ events and a
somewhat smaller number of detected $B^{\pm}\to\pi^{\pm}K_S$ events is
sufficient to reduce the presently allowed region in the ($\alpha,~\gamma$)
plane by  a considerable amount.

\subsection {$\gamma$ from charged $B$ decays: the use of $\eta$ and $\eta'$}

The use of $\eta$ and $\eta'$ allows a determination of $\gamma$ from decays
involving charged $B$ decays alone \cite{ETA}.
When considering final states involving $\eta$ and $\eta'$ one encounters one
additional penguin diagram (a so-called ``vacuum cleaner" diagram),
contributing
to decays involving one or two flavor SU(3) singlet pseudoscalar mesons
\cite{DGR2}. This amplitude ($P_1$) appears in a fixed combination with a
higher-order electroweak penguin contribution in the form $p_1\equiv P_1-
(1/3)P_{EW}$.

Writing the physical states in terms of the SU(3) singlet and
octet states
\beq
\eta=\eta_8\cos\theta - \eta_1\sin\theta,~~~\eta'=\eta_8\sin\theta+\eta_1
\cos\theta,~~~\sin\theta\approx {1\over 3},
\eeq
one finds the following expressions for the four possible $\Delta S=1$
amplitudes of charged $B$ decays to two charmless pseudoscalars:
$$
A(B^+\to \pi^+ K^0)=p'~,~~~~A(B^+\to\pi^0 K^+)={1\over \s}(-p'-t'-c')~,
$$
\beq
A(B^+\to\eta K^+)={1\over\sqrt{3}}(-t'-c'-p'_1)~,~~~~A(B^+\to \eta' K^+)=
{1\over\sqrt{6}}(3p'+t'+c'+4p'_1)~.
\eeq
These amplitudes satisfy a quadrangle relation
$$
\sqrt{6}A(B^+\to \pi^+ K^0) + \sqrt{3}A(B^+\to\pi^0 K^+)
$$
\beq
- 2\sqrt{2}A(B^+\to\eta K^+) - A(B^+\to \eta' K^+) = 0~.
\eeq

A similar quadrangle relation is obeyed by the charge-conjugate amplitudes, and
the relative orientation of the two quadrangles holds information about
weak phases. However, it is clear that each of the two quadrangles cannot be
determined from its four sides given by the measured amplitudes. A closer look
at the expressions of the amplitudes shows that the two quadrangles share a
common base, $A(B^+\to \pi^+ K^0)=A(B^-\to \pi^- \overline{K}^0)$, and the two
sides opposite to the base (involving $\eta$) intersect at a point lying
3/4 of the distance from one vertex to the other. This fixes the shapes of
the quadrangles up to discrete ambiguities. Finally, the phase $\gamma$ can be
determined by relating these amplitudes to that of $B^+\to\pi^+\pi^0$
\beq
\vert A(B^+\to\pi^0 K^+) -  A(B^-\to\pi^0 K^-)\vert =
2{V_{us}\over V_{ud}}{f_K\over f_{\pi}}\vert
A(B^+\to\pi^+\pi^0)\vert\sin\gamma~.
\eeq
Note that in this method we neglect SU(3) breaking terms associated with the
difference betwen creation of nonstrange and strange quark pairs in the final
state of penguin amplitudes.

\section{Beyond the standard model}

\subsection{CP asymmetries and the unitarity triangle}

The above discussion assumes that the only source of CP violation is the phase
of the CKM matrix. Models beyond the SM involve other phases, and consequently
the measurements of CP asymmetries may violate SM constraints on the three
angles of the unitarity triangle \cite{DLN}. Furthermore, even in the absence
of new CP
violating phases, these angles may be affected by new contributions to the
sides of the triangles. The three sides (Fig.~1), $V_{cd}V^*_{cb},
~V_{ud}V^*_{ub}$ and $V_{td}V^*_{tb}$ are measured in $b\to c l\nu,~
b\to u l\nu$ and in $\bd-\bdb$ mixing, respectively.
A variety of models beyond the SM provide new contributions to $\bd-\bdb$ and
$\bs-\bsb$ mixing, but only very rarely do such models involve new
amplitudes which can compete with the $W$-mediated tree-level $b$ decays.
Therefore, whereas two of the sides of the unitarity triangle are usually
stable under new physics effects, the side involving $V_{td}V^*_{tb}$ can be
modified by such effects. In certain models, such as a four generation model
and models involving $Z$-mediated flavor-changing neutral currents (to be
discussed below), the unitarity triangle turns into a quadrangle.

In the phase convention of Eq.(\ref{V})
the three angles $\alpha,~\beta,~\gamma$ are defined as follows:
\beq
\gamma\equiv {\rm Arg}(V_{ud}V^*_{ub})~,~~~~~~ \beta\equiv {\rm Arg}
(V_{tb}V^*_{td})~,~~~~~~\alpha\equiv \pi-\beta-\gamma~.
\eeq
Assuming that new physics affects only $B^0-\overline{B}^0$ mixing, one can
make the
following simple observations about CP asymmetries beyond the SM:
\begin{enumerate}
\item The asymmetry in $B^0_d\to\psi K_S$ measures the phase of $\bd-\bdb$
mixing and is given by $2\beta'$, which in general can be different from
$2\beta$.
\item  The asymmetry in $B^0_d\to\pi^+\pi^-$ measures the phase of $\bd-\bdb$
mixing plus twice the phase of $V^*_{ub}$, and is given by $2\beta'+
2\gamma\equiv 2\pi-2\alpha'$, where $\alpha'\ne\alpha$.
\item The time-dependent rates of $\bs/\bsb\to D^{\pm}_s K^{\mp}$ determine a
phase $\gamma'$ given by the phase of $\bs-\bsb$ mixing plus the phase of
$V^*_{ub}$; in this case $\gamma'\ne\gamma$.
\item The processes $B^{\pm}\to D^0 K^{\pm},~B^{\pm}\to \Dbar K^{\pm},
~B^{\pm}\to D^0_{1(2)} K^{\pm}$ measure the phase of $V^*_{ub}$ given by
$\gamma$.
\end{enumerate}
That is, a measurement of the phase $\gamma$ through the last method will
obey the triangle relation $\alpha'+\beta'+\gamma=\pi$
with the phases of $B^0_d\to\psi K_S$ and $B^0_d\to\pi^+\pi^-$, irrespective of
contributions from new physics. On the other
hand, the phase $\gamma'$ measured by the third method violates this relation.
This demonstrates the importance of measuring phases in a variety of
independent
ways.

Let us note in passing that in certain models, such as multi-Higgs doublet
models with natural flavor conservation (to be discussed below), in spite
of new
contributions to $\bd-\bdb$ and $\bs-\bsb$ mixing, the phases measured in
$B^0_d\to\psi K_S$ and in $\bs/\bsb\to D^{\pm}_s K^{\mp}$ are unaffected,
$\beta'=\beta,~\gamma'=\gamma$. The values measured for these phases may,
however, be inconsistent with the CP conserving measurements of the sides of
the unitarity triangle.

{\subsection {CP asymmetries versus penguin decays}

Models in which CP asymmetries in $B$ decays are affected by new contributions
to $B^0-\overline{B}^0$ mixing will usually also have new amplitudes
contributing
to rare flavor-changing $B$ decays, such as $b\to s X$ and $b\to d X$.
We refer to such processes, involving a photon, a pair of leptons or hadrons
in the final state, as ``penguin" decays.

In the SM both $B^0-\overline{B}^0$ mixing and penguin decays are governed
by the
CKM parameters $V_{ts}$ and $V_{td}$. Unitarity of the CKM matrix implies
\cite{ALON} $\vert V_{ts}/V_{cb}\vert\approx 1$, $0.11 < \vert V_{td}/V_{cb}
\vert < 0.33$, and $\bd-\bdb$ mixing only improves this constraint slightly due
to large hadronic uncertainties, $0.15 < \vert V_{td}/V_{cb}\vert < 0.33$.

The addition of contributions from new physics to $B^0-\overline{B}^0$ mixing
relaxes the above constraints in a model-dependent manner. The new
contributions
depend on new couplings and new mass scales which appear in the models. These
parameters also determine the rate of penguin decays. A recent comprehensive
study \cite{MGDL}, updating previous work, showed that the values of the new
physics parameters, which yield significant effects in $B^0-\overline{B}^0$
mixing, will also lead in a variety of models to large deviations from the SM
predictions for certain penguin decays. Here we wish to briefly summarize
the results of this model-by-model analysis:

\begin{enumerate}
\item {\it Four generations}: The magnitude and phase of $B^0-\overline{B}^0$
mixing
can be substantially changed due to new box-diagram contributions involving
internal $t'$ quarks. For such a region in parameter space, one expects an
order-of-magnitude enhancement (compared to the SM prediction) in the branching
ratio of $\bd\to l^+l^-$ and $B^+\to\phi\pi^+$.
\item {\it $Z$-mediated flavor-changing neutral currents}: The magnitude and
phase of $B^0-\overline{B}^0$ mixing can be altered by a tree-level
$Z$-exchange.
If this effect is large, then the branching ratios of the penguin processes
$b\to s l^+ l^-,~\bs\to l^+\l^-,~\bs\to\phi\pi^0~(b\to d l^+ l^-,~
\bd\to l^+\l^-,~B^+\to\phi\pi^+$) can be enhanced by as much as one (two)
orders-of-magnitude.
\item {\it Multi-Higgs doublet models with natural flavor conservation}:
New box-diagram contributions to $B^0-\overline{B}^0$ mixing with internal
charged
Higgs bosons affect the magnitude of the mixing amplitude but not its phase
(measured, for instance, in $\bd\to\psi K_S$). When this effect is large, the
branching ratios of $\bd,\bs\to l^+ l^-$ are expected to be larger than in
the SM
by up to a factor 5.
\item {\it Multi-Higgs doublet models with flavor-changing neutral scalars}:
Both the magnitude and phase of $B^0-\overline{B}^0$ mixing can be changed
due to a
tree-level exchange of a neutral scalar. In this case one expects no
significant
effects in penguin decays.
\item {\it Left-right symmetric models}: Unless one fine-tunes the right-handed
quark mixing matrix, there are no significant new contributions in
$B^0-\overline{B}^0$ mixing and in penguin $B$ decays.
\item {\it Minimal supersymmetric models}: There are a few new contributions to
$B^0-\overline{B}^0$ mixing, all involving the same phase as in the SM.
Branching
ratios of penguin decays are not changed significantly. However, certain energy
asymmetries, such as the $l^+l^-$ energy asymmetry in $b\to s l^+ l^-$ can be
largely affected.
\item{\it Non-minimal supersymmetric models}: In non-minimal SUSY models with
quark-squark alignment, the SUSY contributions to $B^0-\overline{B}^0$ mixing
and to penguin decays are generally small. In other models, in which all SUSY
parameters are kept free, large contributions with new phases can appear in
$B^0-\overline{B}^0$ mixing and can affect considerably SM predictions for
penguin
decays. However, due to the many parameters involved, such schemes have little
predictivity.
\end{enumerate}

We see that measurements of CP asymmetries and rare penguin decays
give complementary information and can distinguish among the different models.
For instance, in models 3 and 6 one expects $\beta'=\beta,~\gamma'=\gamma$,
whereas in models of type 1, 2 and 4 one has
$\beta'\ne\beta,~\gamma'\ne\gamma$.
In the latter case, one expects different measurements of $\gamma$ in
$B^{\pm}\to D K^{\pm}$ and in $\bs/\bsb\to D_s^{\pm} K^{\mp}$. The three models
1, 2 and 4 can then be distinuished by their different predictions for branching
ratios of penguin decays. To distinguish between models 3 and 6, one
would have to rely on detailed dilepton energy distributions.

\section{Conclusion}

We addressed certain theoretical issues related to three of the main goals of
future $B$ physics experiments, which are:
\begin{itemize}
\item Observation of CP asymmetries in $B^0_d,~B^{\pm},~B^0_s$ decays.
\item Determination of $\alpha,~\beta,~\gamma$ from these asymmetries and
from $B$ decay rates.
\item Detection of deviations from Standard Model asymmetry predictions, which
would provide clues to a more complete theory when combined with information
about rare flavor-changing $B$ decays.
\end{itemize}
\bigskip
\large\noindent
{\bf Acknowledgements}
\normalsize
\bigskip

\noindent
It is a pleasure to thank A. Dighe, O. Hern\'anzez, D. London, J. Rosner and
D. Wyler for very enjoyable collaborations on various topics presented here.
This work was supported in part by the United States-Israel Binational Science
Foundation under Research Grant Agreement 94-00253/2, and by the Fund for
Promotion of Research at the Technion.

\newpage

\def \np#1#2#3{Nucl.~Phys.~B{\bf#1}, #2 (#3)}
\def \plb#1#2#3{Phys.~Lett.~B{\bf#1}, #2 (#3)}
\def \prd#1#2#3{Phys.~Rev.~D {\bf#1}, #2 (#3)}
\def \prl#1#2#3{Phys.~Rev.~Lett.~{\bf#1}, #2 (#3)}
\def \prp#1#2#3{Phys.~Rep.~{\bf#1} #2 (#3)}
\def \ptp#1#2#3{Prog.~Theor.~Phys.~{\bf#1}, #2 (#3)}
\def \rmp#1#2#3{Rev.~Mod.~Phys.~{\bf#1} #2 (#3)}
\def \zpc#1#2#3{Z.~Phys.~C {\bf#1}, #2  (#3)}
\def \ite{{\it et al.}}

\newpage
\hoffset -0.7in

\def\Kbar{\overline{K}^0}
\font\er=cmr10 scaled\magstep0
\def \beq{\begin{equation}}
\def \eeq{\end{equation}}
\def \beqn{\begin{eqnarray}}
\def \eeqn{\end{eqnarray}}
\def \Bbar{\bar B}
\def \Dbar{\bar D}
\def \Kbar{\bar {K}^0}
\def \sf{\hbox{$\scriptstyle{1\over\sqrt2}$}}
\def \st{\sqrt{3}}
\def \sx{\sqrt{6}}
\def \v#1#2{V_{#1#2}}
\def \vc#1#2{V^*_{#1#2}}
\renewcommand{\arraystretch}{1.2}
\begin{table}
\begin{center}
\begin{tabular}{c c c c c c c c c c} \hline
$\Delta S=0$  & \multicolumn{3}{c}{Coefficient of:} &~\vline~&
$\Delta S=1$  & \multicolumn{3}{c}{Coefficient of:}\\
process & $t$ & $c$  & $p$ &~\vline~&
process & $t'$ & $c'$ & $p'$  \\ \hline
$B^+\to\pi^+\pi^0$ & $-1/\sqrt{2}$ & $-1/\sqrt{2}$ & &~\vline~&
$B^+\to\pi^+ K^0$ & & & 1 \\
$B^+\to K^+\Kbar$ &  &  &  $1$ &~\vline~&
$B^+\to\pi^0 K^+$ & $-1/\s$ & $-1/\s$ & $-1/\s$ \\
$B^0_d\to\pi^+\pi^-$ & $-1$ &  &  $-1$ &~\vline~&
$B^0_d\to\pi^-K^+$ & $-1$ & & $-1$ \\
$B^0_d\to\pi^0\pi^0$ &  & $-1/\s$ &  $1/\s$ &~\vline~&
$B^0_d\to\pi^0K^0$ &  & $-1/\s$ & $1/\s$ \\
$B^0_d\to K^+ K^-$ &  &  &  &~\vline~&
$B^0_s\to\pi^+\pi^-$ &  & &  \\
$B^0_d\to K^0 \Kbar$ &  &  & $1$ &~\vline~&
$B^0_s\to\pi^0\pi^0$ &  & &  \\
$B^0_s\to \pi^+ K^-$ & $-1$ &  & $-1$ &~\vline~&
$B^0_s\to K^+ K^-$ & $-1$ & & $-1$ \\
$B^0_s\to \pi^0 \Kbar$ &  & $-1/\s$ & $1/\s$ &~\vline~&
$B^0_s\to K^0 \Kbar$ & & & $1$ \\
\hline
\end{tabular}
\caption{Decomposition of amplitudes for $B$ decays to two light pseudoscalars}
\label{TABLELABEL}
\end{center}
\end{table}
~~
\newpage


\begin{figure}
\caption[]{The CKM unitarity triangle}
\label{name of figure 2}
\end{figure}


\end{document}